\documentstyle[aps,psfig,twocolumn]{revtex}

\begin{document}
%\draft
\topmargin -15mm

%\title{
%Theoretical calculation of uniaxial magnetic anisotropy and magnetostriction in strained CMR films.
%}
\title{
First principles calculation of uniaxial magnetic anisotropy and magnetostriction in strained CMR films
}

\author{A. B. Shick}

\address{Department of Physics, University of California, Davis, CA 95616}

\maketitle

\begin{abstract}

We performed first - principles relativistic full-potential
linearized  augmented plane wave calculations for
strained tetragonal ferromagnetic La(Ba)MnO$_3$ with an assumed experimental
structure of thin strained tetragonal La$_{0.67}$Ca$_{0.33}$MnO$_3$ (LCMO) films
grown on SrTiO$_3$[001] and LaAlO$_3$[001] substrates.
The calculated uniaxial magnetic anisotropy energy (MAE) values,
are in good quantitative agreement with experiment for LCMO films on SrTiO$_3$ substrate.
%For La(Ba)MnO$_3$ on LaAlO$_3$ substrate we calculated the MAE to be positive and confirmed
%the phenomenological prediction.
We also analyze the applicability of linear magnetoelastic theory for describing
the stain dependence of MAE, and estimate magnetostriction coefficient $\lambda_{001}$.

\end{abstract}

%\newpage
\section{Introduction}
%\vspace{0.5cm}

Magnetic multilayers based on ``colossal'' magnetoresistive
(CMR) materials are important for many magnetic applications
including recording media and magnetoresistive sensors.
The particular class of CMR materials, La$_{1-x}$A$_x$MnO$_3$ (A=Sr,Ca,Ba),
is particularly interesting because they share the same basic perovskite
crystal structure with many dielectrics, superconductors, and ferroelectrics.
This stuctural similarity opens interesting possibilities for the epitaxial growth of 
CMR magnetic heterostructures for device applications.

Magnetic anisotropy energy (MAE) and magnetostriction in magnetic CMR heterostructures
is significantly different from the MAE in bulk manganites \cite{exp}.
Uniaxial magnetic anisotropy in thin La$_{0.67}$Ca$_{0.33}$MnO$_3$ (LCMO)
films grown on SrTiO$_3$[001] (STO) substrate  has been measured and interpreted
to be due to the strain arising from a film/substrate lattice mismatch \cite{exp}.
The possibility of producing strained LCMO films grown on a LaAlO$_3$ (LAO) substrate
with the easy magnetization axis along film normal has been proposed.

In order to tackle the origin of the uniaxial MAE in strained CMR films
we performed first - principles calculations of MAE for the
strained tetragonal ferromagnetic La(Ba)MnO$_3$ (L(B)MO) with an assumed experimental
structure of
strained LCMO-films grown on STO and LAO substrates.
We also analyze the applicability of linear magnetoelastic theory for describing
the stain dependence of MAE, and estimate magnetostriction
in LCMO-films.

\section{Computational method}

The experimental data of the MAE and magnetostriction in LCMO
films have been obtained from magnetization curve measurements \cite{exp}.
The 58 nm thick films were grown on [001] oriented STO substrate using
atomic layer by layer molecular-beam epitaxy \cite{Odonnell}.
X-ray diffraction data showed that the films have
a tetragonal unit cell with in-plane lattice
constant 7.3696 a.u. which is consistent with STO substrate lattice constant
and perpendicular lattice constant of 7.2373 a.u. This is fully consistent with
the so-called ``coherent'' regime of the film strain in the ``elastic'' approximation
\cite{vdM} :
film in-plane lattice constant matches with substrate in-plane lattice constant
that results in perpendicular tetragonal strain determined by film Poisson ratio.
As a result there are no misfit dislocations and no thickness dependence of
the perpendicular strain.

An anisotropic energy density of a tetragonal ferromagnetic film on non-magnetic substrate
can be written as \cite{Landau}:
  
\begin{eqnarray}
\label{1}
E/V \; = - K_1^v m_z^2 \; - \; K_2^v m_z^4 \; - \; K_3^v m_x^2 m_y^2 \\ \nonumber
- \frac{2}{t} (K_1^s m_z^2 \; + \; K_2^s m_z^4 \; + \; K_3^s m_x^2 m_y^2) 
\end{eqnarray}
where, $K^v$ terms are 2nd and 4th-order volume-type anisotropy constants,
$K^s$ terms are surface/interface-type 2nd and 4th-order anisotropy constants,
$m_{x,y,z}$ are magnetization cosines with respect to the crystal axes, and
t is a thickness of magnetic film. 
For the particular case of LCMO films considered here $t$ is $\approx$ 150 ML
and surface term in Eq.(1) can be safely neglected. Therefore, 
as a computational model for LCMO film grown on STO and LAO substrates we used 
strained tetragonal bulk La(Ba)MnO$_3$ with the crystal structure parameters chosen
in accordance with experiment \cite{exp}. We also assume that the 4th-order
terms in Eq.(1) are significantly smaller (as usual) than 2nd-order uniaxial
anisotropy constant $K_1^v$, and did not consider them.  

Since the measured films are ferromagnetic at low temperature
for $x=0.33$ Ca consentration, we considered the only ferromagnetic
phase. We also used end-point ferromagnetic LaMnO$_3$ and BaMnO$_3$ 
instead of actual LCMO in spite of the fact that both LMO and BMO
have an antiferromagnetic ground state. The use of ferromagnetic LMO and
BMO to approximate ferromagnetic LCMO films allows us to calculate
the possible range of values for the uniaxial anisotropy in real
LCMO films and to analyze the effect
of electron concentration (changing Mn-site d-occupation) on MAE.

We used a relativistic version \cite{flapwso} of the full-potential
linearized  augmented plane wave method \cite{flapw} to obtain 
the self-consistent solutions of Kohn-Sham-Dirac equations and
ground state charge and spin densities 
for the magnetization directed along [001]-axis.
For both strained LMO and BMO
we used the special k-point method \cite{BZ} for the
Brillouin zone (BZ) integrations. 348 k-points mesh in the 1/8th irreducible
part of the BZ \cite{symmetry} is employed for
the self-consistent calculations
and a Gaussian
broadening ($\sigma = 0.0014 \;Ry$) is used for the eigenstates weighting.
Lattice harmonics with angular momentum $l$ up to 8
are used to expand the charge and spin densities and wavefunctions
within the muffin - tin sphere.
More than 100 plane waves per atom/spin are used as the first
variational basis set to solve the scalar-relativistic Kohn-Sham equations;
all occupied and empty states up to 2 Ry above $E_F$ are used
as a second variational basis set to calculate spin-orbit coupling matrix elements.
Self-consistency is achieved to within  $1\times10^{-5} e/(a.u.)^3$
for charge and spin densities and to within $2\times10^{-5} Ry$
for the all-electron total energy.

The spin and orbital magnetic moments for magnetization along the [001] axis
for strained LMO and BMO
for STO and LAO substrates are shown in Table 1. The orbital magnetic moments are antiparallel
to spin moments, which is consistent with the atomic third Hund's rule for the case of
less than half-filled d-shell. There is a decrease of spin moment and an increase of absolute value of the
orbital moment with substitution of La by Ba due to decrease of Mn d-occupation. Assuming 
single-site approximation for the magnetic moment we can estimate LCMO spin moment to be 2.85 $\mu_B$ and
orbital moment -0.011 $\mu_B$ for the STO substrate and 2.77 and -0.014 $\mu_B$ respectively
for LAO substrate.

The MAE \cite{MAE} is then obtained by applying the force theorem to
the spin - axis rotation \cite{Igor}:
from the self-consistent ground state charge and spin density obtained
for the [001] spin axis,
a calculation of the band structure for [100] spin axis orientation is performed;
difference of the single particle eigenvalue sums is then taken
to be the MAE. 
For the MAE calculations we used special k-points mesh 
which was chosen with respect to the magnetic symmetry for the [100]
spin axis \cite{Weinberger}. Since the local force theorem was used, the symmetry of
eigenvalues rather than Hamiltonian was considered: for the tetragonal
symmetry and magnetization directed along [100] it leads to exclusion of fourfold rotations
with respect to [001] axis and leaves only eight space group symmetry operations
(mmm space group) which are used to generate a set of 
irreducible k-points in 1/8th of BZ (different from those for self-consistent
caculations).

To achieve convergent results for the MAE, we have increased
the number of
the BZ k-points and have done the band calculation for both the [001] and [100]
spin axes using the ground state charge and spin densities.
The MAE dependence on the number of k-points for the LMO(STO) is shown in Fig. 1.
Surprisingly good convergence (less than 1 $\mu eV$) was achieved for 
1000 k-points in 1/8th of BZ (8000 in full BZ). This set of k-points was then
used for all MAE calculations.

\section{Results and discussion}
%\section{results}
The calculated uniaxial MAE
for all four cases (L(B)MO(STO) and L(B)MO(LAO)) are shown in Table II.
The calculated MAE values are
in very good agreement with experiment for LCMO films ($-56 \; \mu eV$)
on a STO substrate. This is direct numerical evidence of the magnetoelastic origin
of the uniaxial MAE in strained LCMO films and supports quantitatively an interpretation
of experiment given in Ref. \cite{exp}.
For LMO(LAO) and BMO(LAO) we calculated the MAE to be positive.
It is in qualitative agreement with very recent experiments for La$_{0.8}$Ca$_{0.2}$MnO$_3$ films
grown on LAO substrate \cite{exp2}.

%and thereby confirmed the prediction of Ref. \cite{exp}.
In addition to calculated magnetocrystalline MAE caused by
spin-orbit interaction 
we have to take into account demagnetization energy (DE) which is
due to ``classic'' magnetic dipole-dipole interaction. For the ferromagnetic
film the DE can be evaluated numerically \cite{Weinberger2}.
%and is shown
%to be mainly due-to dipole-dipole interaction of the magnetic moments
%in the same magnetic plane.
We calculated the DE using the results of Ref. \cite{Weinberger2} and
taking into account that magnetic moment
in L(B)MO is located at Mn-site (cf. Table I.). 

The values of DE for  
all four cases (L(B)MO(STO) and L(B)MO(LAO)) are shown in Table II.
(We took into account the only in-plane dipole-dipole interaction between
Mn-sites.)
For the case of L(B)MO(STO) both MAE and DE keep magnetization
in the film plane. For the case of LMO(STO) the sum of MAE and
DE is negative ($-28 \; \mu eV$) and magnetization
is in the film plane. For the BMO(STO) the total uniaxial MAE
is positive ($42 \; \mu eV$) and magnetization is 
directed along the film normal.
 
Assuming 
single-site approximation for the  magnetic anisotropy
we can estimate total uniaxial MAE $\approx -7 \; \mu eV \; per \; cell$ for LCMO film on
LAO substrate. It means that the absolute value of
the magnetoelastic MAE is very close to the absolute value of the demagnetization field
for LCMO film on LAO substrate.

%\section{model}
%We also analyze the applicability of linear magnetoelastic theory for describing
%the stain dependence of MAE, and estimate magnetostriction \cite{flapwso}
%in LCMO-films.

To check the applicability of linear magnetoelastic theory we map our results
onto the usual anisotropic magnetoelastic energy dependence on strain:

\begin{equation}
\label{eq:MAE}
E \; =  B_1 \; (e_{z} - e_{0}) \; m_z^2 + const \, ,
\end{equation}

\noindent where $e_0$ is a biaxial (in-plane) strain due to the film/substrate lattice mismatch,
$e_z$ is a uniaxial (along [001]) strain, $B_1$ is a magnetoelastic coefficient,
and $m_z$ is a magnetization cosine with respect to the z([001])-axis.

Using Eq.(\ref{eq:MAE}), the values of strain \cite{strain} and MAE, we calculate the values 
of $B_1$ for L(B)MO for STO and LAO substrates (cf. Tab. II). There is quite pronounced
variation of $B_1$ for LMO for the different substrates. It means that for LMO there is
significant deviation from linear MAE dependence on strain (Eq.(\ref{eq:MAE})), and linear
theory is rather qualitative. For the case of BMO the variation of $B_1$ is smaller,
therefore linear theory is more reliable. Note, that calculated values of magnetoelasic
coefficient $B_1$ are in very reasonable agreement with the experimentally derived value
$-6.7 \times 10^7 \; erg/cm^3 \; (-2.44 \; meV \; per \; cell)$ for LCMO films on STO substrate.

In spite of the fact that linear theory
gives the only a semi-quantitative description of uniaxial MAE dependence on strain,
our calculations show that one can still use it for estimation of the sign and order of magnitude of the
magnetostriction coefficient $\lambda_{001}$. To determine $\lambda_{001}$ one has to minimize
the sum of Eq.(\ref{eq:MAE}) and elastic energy \cite{clark}:

\begin{eqnarray}
\label{eq:el}
E_{elastic} &=& \frac{1}{2}c_{11}e_{z}^2\;+\;2c_{12}e_{z}e_{0} + const \, ,
\end{eqnarray}
  
\noindent where, $c_{11}$ and $c_{12}$ are elastic moduli, with respect to $e_z$, with $e_0$ fixed by
substrate. It leads to the following expression for the magnetostriction constant $\lambda_{001}$
(see for details Ref. \cite{flapwso}):

\begin{equation}
\label{eq:mscoeff}
\lambda_{001} \; =  \; - \;\frac{2}{3}\;\frac{B_{1}}{c_{11}}
\end{equation}

Assuming $c_{11} = 2 \times 10^{12} \; erg/cm^3$ \cite{exp} and using calculated values
of $B_1$ (cf. Tab. II)
we calculate $\lambda_{001}$ for L(B)MO films on STO and LAO substrates (cf. Table III).
Our calculated values of $\lambda_{001}$ agree in sign and order of magnitude with
the experimentally derived value of $7 \times 10^{-5} $. However, we have to note that 
it is unclear what definition for $\lambda_{001}$ in terms of $B_1$ and $c_{11}$
was used in Ref. \cite{exp}: using the data of Ref. \cite{exp} and Eq.(\ref{eq:mscoeff})
we calculated $\lambda_{001}= 2.23 \times 10^{-5}$ in very good agreement with our results.  

\section{Conclusion}

To summarize, 
we have shown quantitatively that the observed uniaxial anisotropy in LCMO films grown on
STO substrate is caused by strain arising from film/substrate lattice constant mismatch.
It is also shown that the magnetoelastic MAE is positive for LCMO films on LAO substrate
and absolute value of MAE is close to the absolute value of demagnetization energy.
By varying the strain arising 
from a film/substrate lattice constant mismatch 
one can control the uniaxial anizotropy for LCMO films in order to
overcome the demagnetization energy and provide [001] 
spontaneous magnetization direction.  

I am grateful to W. E. Pickett, A. J. Freeman and O. N. Mryasov for helpful
discussion. This research was supported by the National Science foundation Grant
DMR-9802076.

\newpage

%TABLE:
\begin{table}
\caption
{Experimental lattice constant (a, in a.u.) and c/a ratio for tetragonal strained LCMO films 
on STO and LAO substrates; spin ($M_s$) and orbital ($M_l$) magnetic moments for LMO and BMO
for the [001] spin direction.} 
\begin{tabular}{cccccc}
Substrate: & STO         &       & LAO &\\
a=b        & 7.3697      &       & 7.1618      \\
c/a        & 0.982       &       & 1.047 \\
Film:      & LMO         &  BMO  & LMO   & BMO\\
$M_s$ (Mn) & 2.996       & 2.563 & 2.880 & 2.542 \\
$M_s$ (Total) & 3.145    & 2.727 & 3.022 & 2.712 \\
$M_l$ (Mn) & -0.004      & -0.023& -0.007& -0.023 \\
$M_l$ (Total) & -0.002   & -0.021& -0.006& -0.024 \\
\end{tabular}
\end{table}

\begin{table}
\caption
{Biaxial ($e_0$) and uniaxial ($e_z$) strains, uniaxial MAE ($\mu eV$),
demagnetization energy (DE) ($\mu eV$), 
$B_1$ ($meV$)  
for LMO and BMO films on STO and LAO substrates.}
\begin{tabular}{cccccc}
Substrate: & STO         &       & LAO &\\
$e_0$      &  0.008      &       & -0.021      \\
$e_z$      & -0.0101     &       & 0.0258      \\
Film:      & LMO         &  BMO  & LMO   & BMO\\
MAE        & -40.9       & -53.3 & 37.7  & 100.2 \\
DE         & -65.7       & -48.7 & -66.1 & -58.3 \\
$B_1$      & -2.259      & -2.944& -0.806& -2.141\\
\end{tabular}
\end{table}

\begin{table}
\caption
{$\lambda_{001} \times 10^5 $  
for LMO and BMO films on STO and LAO substrates.}
\begin{tabular}{cccccc}
Substrate:     & STO         &       & LAO &\\
Film:          & LMO         &  BMO  & LMO   & BMO\\
$\lambda_{001}$& 2.1         &  2.74 & 0.76 & 2.01   \\
\end{tabular}
\end{table}

%\newpage
%\centerline{FIGURES:}

%Figure 1.: The calculated uniaxial MAE dependence on k-points number in 1/8th
%irreducible part of the BZ.

\begin{center}
\begin{figure}
\label{mae}
\psfig{file=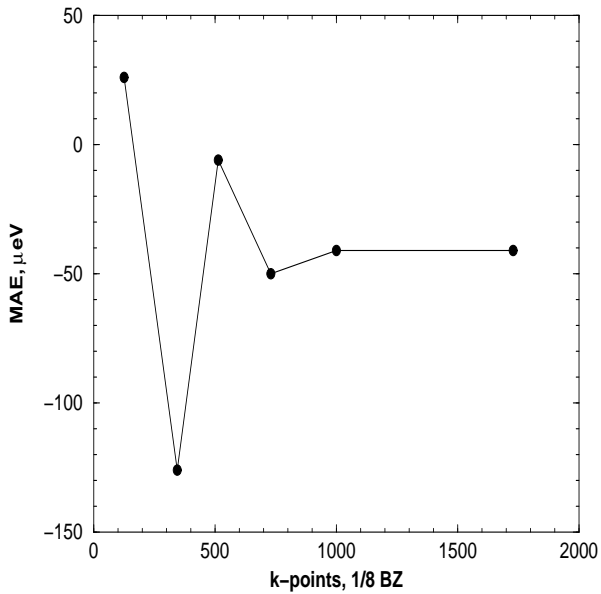,width=8cm,height=8cm}
\caption{The calculated uniaxial MAE dependence on k-points number in 1/8th
irreducible part of the BZ.}
\end{figure}
\end{center}


\begin{references}
\bibitem{symmetry} The space group was chosen to be consistent with the proper
magnetic symmetry in the presence of spin-orbit interaction. For tetragonal
symmetry and magnetization directed along [001] it leads to exclusion of mirror
planes parallel to [001] and leaves only eight space group symmetry operations
(4/m space group).
\bibitem{MAE} MAE is defined as an energy difference for the 
(100) and (001) magnetization directions.
\bibitem{strain} We are accounting strain from cubic ``LCMO'' with a=b=c=7.3112 a.u.
in accordance to Ref. \cite{exp}.
\end{references}
\end{document}